\documentclass[sigconf]{acmart}
\usepackage{threeparttable}

\AtBeginDocument{%
  }

\setcopyright{acmcopyright}
\copyrightyear{2018}
\acmYear{2018}
\acmDOI{XXXXXXX.XXXXXXX}

\acmConference[Conference acronym 'XX]{Make sure to enter the correct
  conference title from your rights confirmation emai}{June 03--05,
  2018}{Woodstock, NY}
\acmPrice{15.00}
\acmISBN{978-1-4503-XXXX-X/18/06}

\copyrightyear{2023} 
\acmYear{2023} 
\setcopyright{acmlicensed}\acmConference[SIGIR '23]{Proceedings of the 46th International ACM SIGIR Conference on Research and Development in Information Retrieval}{July 23--27, 2023}{Taipei, Taiwan}
\acmBooktitle{Proceedings of the 46th International ACM SIGIR Conference on Research and Development in Information Retrieval (SIGIR '23), July 23--27, 2023, Taipei, Taiwan}
\acmPrice{15.00}
\acmDOI{10.1145/XXXXXXX.XXXXXXX}
\acmISBN{978-1-4503-XXXX-X/18/06}

\begin{document}

\title{Curriculum Modeling the Dependence among Targets with Multi-task Learning for Financial Marketing}

\author{Yunpeng Weng}
\authornotemark[1]
\email{edwinweng@tencent.com}
\orcid{0000-0001-7593-2169}
\affiliation{%
  \institution{FiT, Tencent}
  \city{Shenzhen}
  \state{Guangdong}
  \country{China}
  \postcode{518057}
}

\author{Xing Tang}
\authornote{Contributed equally}
\email{shawntang@tencent.com}
\orcid{0000-0003-4360-0754}
\affiliation{%
  \institution{FiT, Tencent}
  \city{Shenzhen}
  \state{Guangdong}
  \country{China}
  \postcode{518057}
}

\author{Liang Chen}
\authornotemark[2]
\email{leoncuhk@gmail.com}
\orcid{0000-0002-3149-0239}
\affiliation{%
  \institution{FiT, Tencent}
  \city{Shenzhen}
  \state{Guangdong}
  \country{China}
  \postcode{518057}
}

\author{Xiuqiang He}
\authornote{Corresponding authors}
\email{xiuqianghe@tencent.com}
\orcid{0000-0002-4115-8205}
\affiliation{%
  \institution{FiT, Tencent}
  \city{Shenzhen}
  \state{Guangdong}
  \country{China}
  \postcode{518057}
}

\renewcommand{\shortauthors}{Weng et al.}

\begin{abstract}
Multi-task learning for various real-world applications usually involves tasks with logical sequential dependence. For example, in online marketing, the cascade behavior pattern of $impression \rightarrow click \rightarrow conversion$ is usually modeled as multiple tasks in a multi-task manner, where the sequential dependence between tasks is simply connected with an explicitly defined function or implicitly transferred information in current works. These methods alleviate the data sparsity problem for long-path sequential tasks as the positive feedback becomes sparser along with the task sequence. However, the error accumulation and negative transfer will be a severe problem for downstream tasks. Especially, at the beginning stage of training, the optimization for parameters of former tasks is not converged yet, and thus the information transferred to downstream tasks is negative. In this paper, we propose a prior information merged model (\textbf{PIMM}), which explicitly models the logical dependence among tasks with a novel prior information merged (\textbf{PIM}) module for multiple sequential dependence task learning in a curriculum manner. Specifically, the PIM randomly selects the true label information or the prior task prediction with a soft sampling strategy to transfer to the downstream task during the training. Following an easy-to-difficult curriculum paradigm, we dynamically adjust the sampling probability to ensure that the downstream task will get the effective information along with the training. The offline experimental results on both public and product datasets verify that PIMM outperforms state-of-the-art baselines. Moreover, we deploy the PIMM in a large-scale FinTech platform, and the online experiments also demonstrate the effectiveness of PIMM.

\end{abstract}





\maketitle

\section{Introduction}
Multi-task learning (MTL) has been widely introduced in many real-world applications such as product recommendation and online marketing \cite{zhao2019recommending,li2020multi,bai2022contrastive,xi2021modeling,zhu2021learning}. 
An MTL model aims to learn multiple related tasks simultaneously and leverages the shared representation to improve performance for all the tasks\cite{crawshaw2020multi}. In these applications, some tasks are logically dependent on each other, which indicates that downstream task requires information from prior tasks. As a result, we should exploit the knowledge about such a dependency relationship and help the MTL model more effectively share the learned information among tasks. In all, modeling the relationships among tasks properly is one of the critical issues in MTL \cite{zhang2021survey}.

\begin{figure}[!t]
  \centering
  \includegraphics[width=0.65\linewidth]{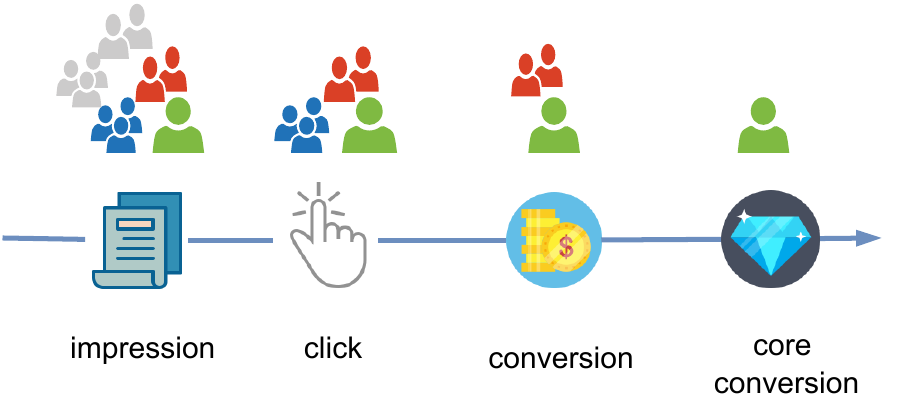}
  \vspace{-0.4cm}
  \caption{The customer conversion path for a financial marketing campaign. Core conversion (purchase over 1,000 CNY) is the primary target due to the commercial consideration.}
  \Description{pim architecture}
  \label{path}
\end{figure}

For many real-world commercial applications, customer conversion follows a sequential behavior pattern. For instance, when distributing mutual funds to customers online, we usually conduct some marketing campaigns to promote revenue. As Figure \ref{path} illustrates, the user behavior in a campaign follows the path \textit{impression} $\rightarrow$ \textit{click} $\rightarrow$ \textit{conversion} $\rightarrow$ \textit{core conversion}, which indicates that only if the positive feedback in the former task occurs, the positive feedback can be expected in the subsequent task, i.e. the target task \textit{core conversion} is progressive. The data sparsity problem thereby becomes severe along with the behavior path. A common way to handle this problem is to formulate it as a multi-task learning regime. On the one hand, notice that core conversion might occur only after the click and purchase behavior occurred. Hence, the regime requires taking the logical dependencies between tasks into account. On the other hand, the downstream tasks will be affected by the former task at the early stage of training, which leads to a phenomenon called \textit{error accumulation}. Specifically, the downstream tasks will make predictions based on the predicted label of former tasks. However, when the optimization is not converged, the performance of the target task will suffer greatly from the accumulated loss of former tasks.

Previous efforts \cite{ma2018modeling,tang2020progressive,qin2020multitask} have been made to share information among tasks, which can alleviate the data sparse problem.
Nevertheless, they ignore modeling the sequential logical dependence among tasks\cite{xi2021modeling}. In recent years, the research on MTL with task dependency has received increasing attention. A collection of models capture the task dependency by simply considering the explicit estimated probability of the prior task with a defined function. Ma et al. \cite{ma2018entire} proposed the Entire Space Multi-task Model (ESMM), which explicitly transfers the predicted probability of the prior task by multiplying it with an estimated conditional probability. Furthermore, wen et al. \cite{wen2020entire} consider more auxiliary tasks in their proposed $ESM^2$ for probability transferring. However, the simple function cannot encode enough information among tasks. To enhance the shared information, several methods focus on exploiting the implicit prior task representation. Adaptive Information Transfer Multi-task (AITM) framework \cite{xi2021modeling} transfers implicit information from the former task's hidden layer in the corresponding tower with the self-attention mechanism. Inspired by AITM, Wu \cite{wu2022mncm} proposed a Multi-level Network Cascades Model (MNCM). MNCM replaces the shared-bottom structure of AITM with the experts-bottom structure and uses Expert-Level Information Transfer Module (EITM) to transfer implicit representation among task-specific experts. However, the prior task information is not fully utilized since they ignore the explicit logical dependency information of the previous task. Besides, the aforementioned methods also lead to error accumulation and negative transfer for downstream tasks.  Especially, at the beginning stage of training, the optimization for parameters of former tasks is not converged yet, and thus the task-specific information transferred to downstream tasks could be misleading, which could exacerbate the difficulty of target tasks learning as they are usually at the end of the path.

To tackle these challenges, we proposed a prior information merged model (\textbf{PIMM}) in this paper, which explicitly models the logical dependence among tasks with a prior information merged (\textbf{PIM}) module for multiple sequential dependence task learning in a curriculum manner. PIMM leverages not only the implicit representation but also the explicit premise information. The implicit representation contains rich task-specific knowledge learned in the tower network of the former task. The explicit premise information indicates the probability that the prior positive feedback occurs. Besides, we introduce a curriculum modeling manner to facilitate information transfer between multiple tasks. Specifically, the PIM randomly selects the true label information or the prior task prediction as explicit premise information. Notice that we introduce a soft sampling strategy following an easy-to-difficult curriculum paradigm to overcome the error accumulation phenomenon. To demonstrate the superiority of our proposed PIMM, we conduct extensive experiments on a publicly available dataset collected from traffic logs of Taobao’s recommender system \footnote{https://tianchi.aliyun.com/dataset/408} and on a real-world industrial dataset collected from the customer acquisition campaigns in a large-scale FinTech platform.  Furthermore, we also conduct online experiments to verify the effectiveness of PIMM in a real-world financial customer acquisition scenario.

\section{The proposed approach}

\subsection{Problem Formulation}
In this paper, we focus on the problem of modeling a collection of binary classification tasks $T=\{T_0, T_1,..., T_M\}$ with sequential dependence.  The corresponding labels of each training sample are $Y=\{y_1, y_2,...,y_M\}$ where M is the number of tasks.  Only if $y_i = 1$, the latter task's label $y_{i+1}$ might be 1, which means only the following three situations are valid for any two adjacent tasks: (1)\ $y_i=0 \& y_{i+1}=0$, (2)\ $y_i=1 \& y_{i+1}=0$, (3)\ $y_i=1 \& y_{i+1}=1$.

\subsection{The Structure of PIMM}

\begin{figure}[htbp]
  \centering
  \includegraphics[width=0.80\linewidth]{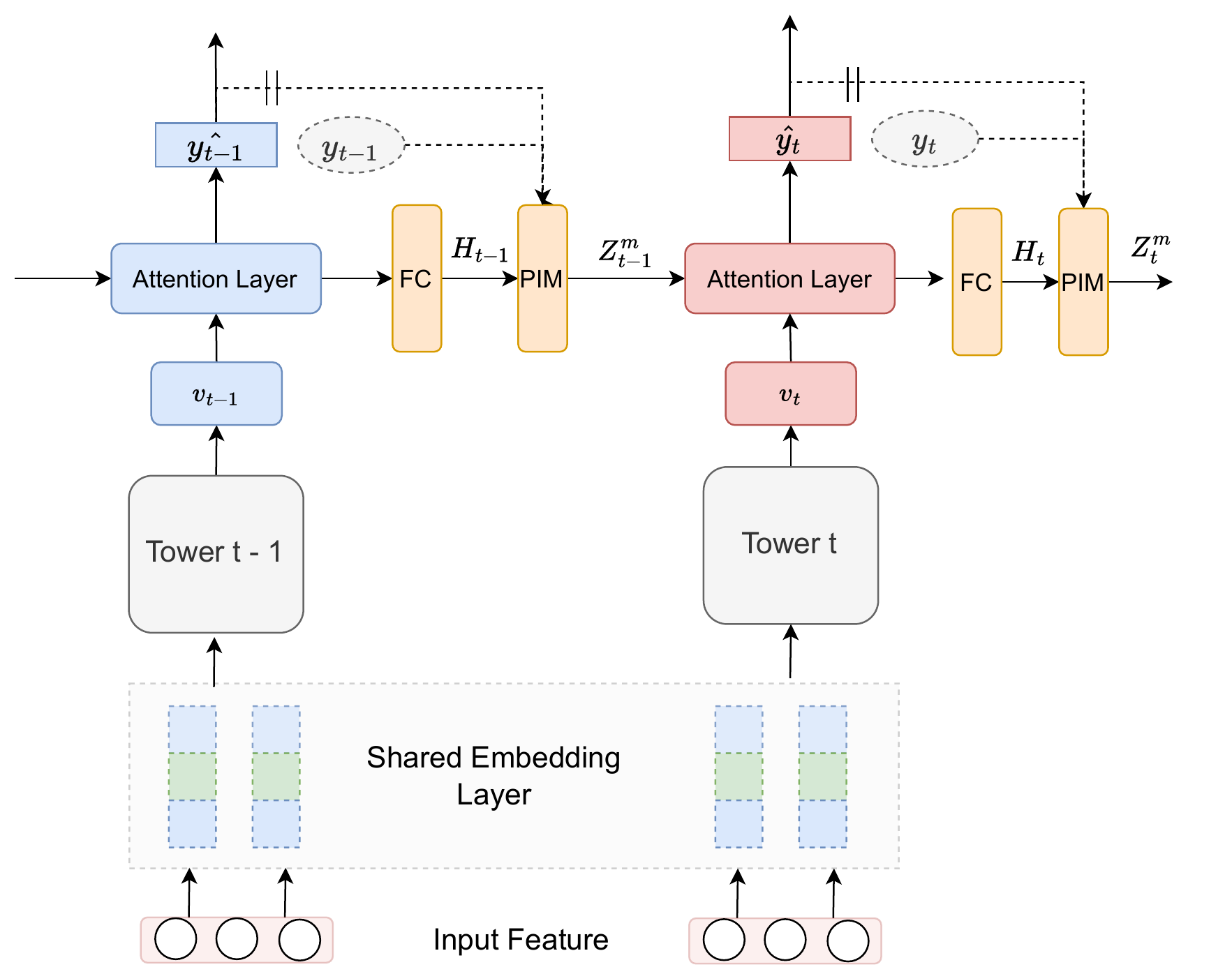}
  \vspace{-0.25cm}
  \caption{
The architecture of PIMM. FC represents a fully-connected layer. "$\Vert$" represents the gradient that is truncated.}
  \Description{pimm architecture}
  \label{pimm}
\end{figure}

We firstly present the overview structure in Figure \ref{pimm}, each feature field $x_i$ will be embedded as a low-dimension vector in the shared embedding layer. Subsequently, all output field vectors are concatenated and fed to all task-specific towers. Note that the backbone of PIMM could be easily extended to more complex structures such as expert-bottom structures. Besides, the advanced models for representation learning can also be utilized as the bottom or tower networks such as graph neural networks\cite{wu2022graph,10.1145/3442201}, feature interaction models\cite{guo2017deepfm,lian2018xdeepfm,9835208}, and so on. In this paper, we focus on the problem of representing and utilizing information about logical dependence among tasks. For all tasks except the first one, we use the prior information merged (PIM) module to obtain the transfer information. The PIM first generates an embedding containing explicit premise information from ground truth or the estimated prediction of the former task with a mutable probability. Then PIM merges the explicit information with the transferred hidden vector that contains implicit information with an addition operation. The detail of PIM will be described in Section \ref{section_pim}. 
Given the output of \textit{t-th} task-specific tower network $v_t \in \mathcal{R}^d$ and the output of PIM $Z^{m}_{t-1}$ which is the prior information merged representation contains the knowledge about the former task, we employ a self-attention layer with the residual connection to generate the representation $U_t$ for the prediction of task $t$:
\begin{equation}
U_t = v_t + \sum_{ \textbf{a} \in \{v_{t}, Z^{m}_{t-1}\}}{ w_a \cdot V(\textbf{a})} 
\end{equation}

where $V$ is the linear transformation function that projects $v_t$ to a new d-dimension vector. $w_a$ is the attention weight, which is calculated as:
\begin{equation}
w_a^{*} = \frac{Q(\textbf{a}) \cdot K(\textbf{a})}{\sqrt{d}}  
\end{equation}

\begin{equation}
w_a = softmax(w_a^{*}) = \frac{\exp(w^{*}_a)}{\sum_{\textbf{a}}{\exp(w^{*}_a)}}
\end{equation}
where Q and K are linear transformation that project $\textbf{a} \in \{v_{t}, Z^{m}_{t-1}\}$ to d-dimension vectors as queries and keys respectively.

Besides being fed to the output layer of task t, $U_t$ will be utilized to generate the implicit representation $H_t$ with a fully-connected layer and be transferred to the subsequent PIM module. Specially,  there is no attention layer for the first task since it has no former task. Therefore, for the first task, $v_0$ will be directly fed to the output layer and $H_0$ is $FC(v_0)$.

\subsection{Prior Information Merged Module}\label{section_pim}
\begin{figure}[!t]
  \centering
  \includegraphics[width=0.75\linewidth]{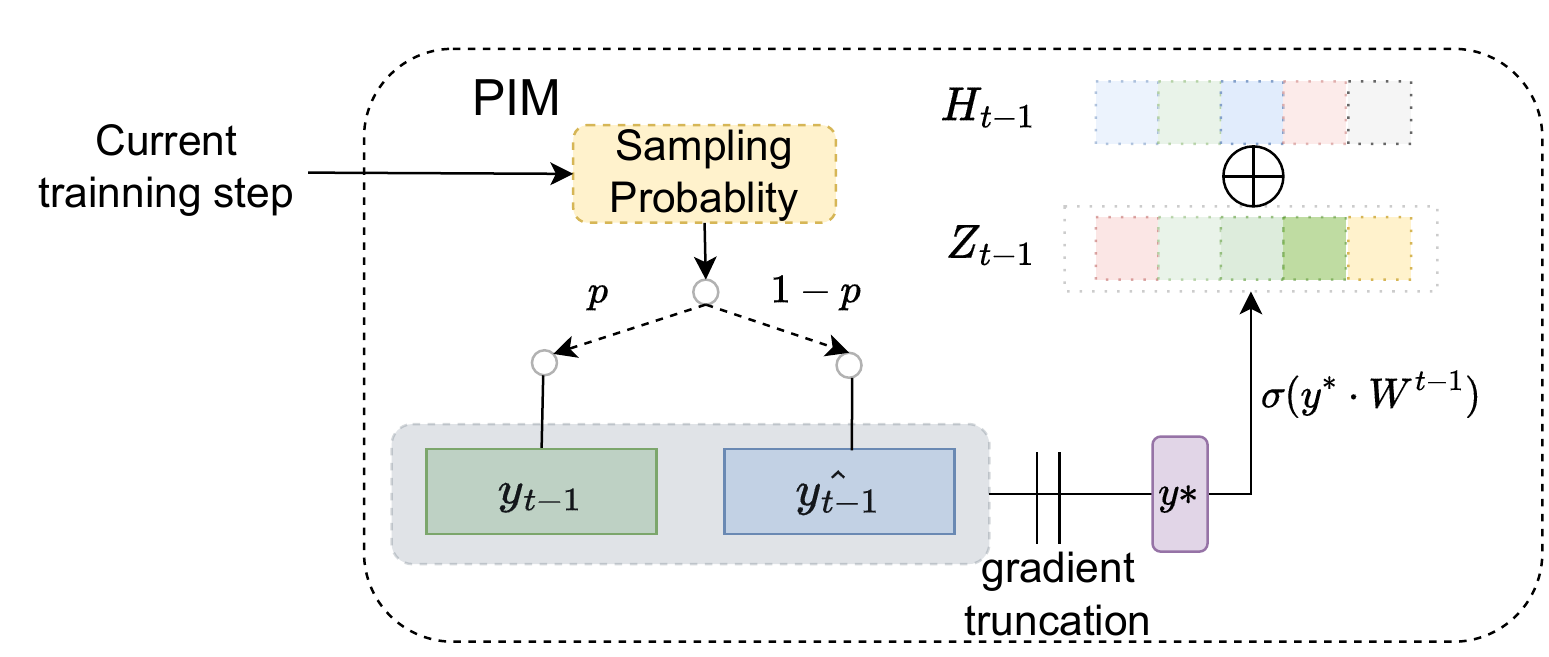}
  \caption{Prior Information Merged (PIM) Module.}
  \Description{pim architecture}
  \label{pim_overview}
\end{figure}
To leverage the explicit premise information from the former task, a naive idea is directly transferring the predicted result of $(t-1)$-th task. However, for some applications such as multi-step conversion prediction in the FinTech platform, the path of sequential dependence targets is long and the error of estimated probability will be transferred accumulatively. Especially at the early stage of training, the optimization for parameters of previous tasks is not converged yet and the predictions are not accurate enough. Inspired by the concept of Curriculum learning\cite{wang2021survey}, we implement an easy-to-difficult training strategy in the Prior Information Merged (PIM) Module as shown in Figure \ref{pim_overview}. The PIM first calculates the sampling probability $p$ according to the current training step. For each training sample in the batch, PIM randomly selects the label $y_{t-1}$ or the predicted probability $\hat{y_{t-1}}$ of the former task with the sampling probability of $p$ and $1-p$ respectively. The sampling probability is mutable during the training process. At the beginning of training, $p$ is a large value and gradually decreases as the training process advances until $p$ reaches the set lower limit. The sampling probability $p$ can be formulated as:
\begin{equation}
p = \max\{\alpha - E \times S, \beta\}
\end{equation}
where $\alpha$ is the initial value of $p$. $E = 0,1,2,...,K$ is the current training epoch. $S$ is the decrease speed and $\beta \ge 0$ is the set lower limit.  At inference, only the predicted probability $\hat{y_{t-1}}$ can be utilized.  In this way, the PIM tends to select the ground truth when the model is not well-trained to avoid transferring misleading prior information in the beginning. As the training advances and the predictions of the model becomes accurate,  PIM is more likely to choose the predicted results gradually and thus eliminates the gap between training and inference.  The selected explicit premise information $y^{*} \in \{y_{t-1}, \hat{y_{t-1}}\}$ is then transformed into a d-dimension vector $Z_{t-1}$ with the following operation:
\begin{equation}
Z_{t-1} = \sigma ( y^{*} \cdot W^{t-1} )
\end{equation}
where $W^{t-1} \in \mathbb{R}^{1 \times d}$ is a trainable variable. To avoid affecting the upstream tasks learning, we apply the gradient truncation operation on the selected $y^*$.
Finally, we obtain the prior information merged task-specific knowledge $Z^{m}_{t-1}$ by adding the transferred implicit representation $H_{t-1}$ from the former task and $Z_{t-1}$:
\begin{equation}
Z^{m}_{t-1} = Z_{t-1} + H_{t-1}
\end{equation}
The curriculum sequential task modeling strategy makes PIMM focus on task-specific parameters learning with the accurate prior information in the beginning and gradually increases the generality as the sampling probability $p$ decreases. 
\section{Experiments}
\subsection{Datasets and Evaluation Metrics}
In the offline experiments, we compare our proposed PIMM model with several baselines including some MTL models on a publicly available dataset and an industrial private dataset:

\textbf{The Ali-CCP dataset}: Alibaba Click and Conversion Prediction dataset\cite{ma2018entire} consists of over 42 million samples in the training set and over 43 million samples in the testing set with sequential labels of click and purchase. Following the previous research \cite{xi2021modeling}, we use the single-values categorical features and randomly retain 10\% samples from the training set as a validation set.

\textbf{The industrial dataset}: We collect logs from historical campaigns from a FinTech platform. The training set contains about 3 million records sampled from 2022/09/01 to 2023/01/04 and the testing set contains 5.8 million records collected after the training period. The dataset contains three sequential dependence tasks: click $\rightarrow$ conversion $\rightarrow$ core conversion, where the definition of core conversion is to invest over 1 thousand CNY in mutual funds distributed in the platform.  

For all models, we report the average AUC and standard deviation (std) over five runs with different random seeds to reduce randomness.

\subsection{Settings}
The baseline models are listed as follows: 

\textbf{Shared-Bottom}\cite{caruana1997multitask}: The Shared-Bottom is a simple MTL structure. All tasks share the same bottom layers and task-specific towers are employed for each task individually. 

\textbf{MMoE}\cite{ma2018modeling}: The MMoE uses multiple expert networks as the bottom structure and applies the gate mechanism to integrate the expert outputs for different tasks. 

\textbf{ESMM}\cite{ma2018entire}: The ESMM transfers the output probability from the previous task $p_{t-1}$ and multiplies it with the estimated conditional probability $p_{t}$ to model the sequential dependence. 

\textbf{PLE}\cite{tang2020progressive}: The PLE introduces task-specific experts besides the shared experts to help mitigate the seesaw phenomenon in MTL. 

\textbf{AITM}\cite{xi2021modeling}: The AITM model adaptively transfers implicit representation from the former task with the attention mechanism. 

\textbf{MNCM}\cite{wu2022mncm}: The MNCM improves the AITM model by utilizing the task-specific and shared experts pattern.

Following the experimental setting of AITM \cite{xi2021modeling}, the embedding size is set to be 5 and 24 for the experiments on the Ali-CCP dataset and industrial dataset respectively. All models are trained with Adam \cite{kingma2014adam} optimizer. We sweep over the learning rate in \{0.0005, 0.001, 0.0015,0.002\} and select the best-performance parameter for all models. For the models with shared-bottom structures, the hidden layer sizes of each task-specific tower are [128,64,32] for the Ali-CCP dataset and [256,128,64] for the industrial dataset. For MNCM and PLE, the number of shared experts is 2 and each task has two task-specific experts. For the MMoE model, the share experts number are 4 and 6 for the Ali-CCP dataset and the industrial dataset respectively. We set $\alpha = 0.5$, $S = 0.25$ , $\beta = 0.25$ for PIMM on the Ali-CCP dataset and  $\alpha = \frac{2}{3} $, $S = \frac{1}{3} $ , $\beta = 0$ on the industrial dataset respectively.

\subsection{Results}

\begin{table}
  \caption{The AUC performance (mean $\pm$ std) of different models on the Ali-CCP dataset.}
  \label{table1}
  
\begin{threeparttable}
  \begin{tabular}{ccc}
    \toprule
    Model & click  & purchase \\
    \midrule
    shared-bottom & 0.6068 $\pm$ 0.0008 & 0.6388 $\pm$ 0.0068 \\
    ESMM & 0.6060 $\pm$ 0.0012 & 0.6417 $\pm$ 0.0064\\
    MMoE & 0.6072 $\pm$ 0.0006 & 0.6463 $\pm$ 0.0062\\
    PLE &  0.6059 $\pm$ 0.0018 & 0.6468 $\pm$ 0.0086\\
    AITM &  0.6072 $\pm$ 0.0013 & 0.6508 $\pm$ 0.0028\\
    MNCM &  0.6071 $\pm$ 0.0011 & 0.6511 $\pm$ 0.0032\\
    PIMM &  $\textbf{0.6075 $\pm$ 0.0010}$ & $\textbf{0.6561 $\pm$ 0.0030}^{\ddagger}$\\
  \bottomrule
\end{tabular}
      \begin{tablenotes}
        \footnotesize
        \item[$\ddagger$] indicates statistically significant improvement over the best baseline(p-value < 0.05).
      \end{tablenotes}
\end{threeparttable}

\end{table}
\begin{table}
  \caption{The AUC performance (mean $\pm$ std) of different models on the industrial dataset.}
  \label{table2}
\scalebox{0.83}{
\begin{threeparttable}
  \begin{tabular}{cccc}
    \toprule
    Model & click  & conversion  & core conversion\\
    \midrule
    shared-bottom & 0.7397 $\pm$ 0.0051 &  0.7106 $\pm$ 0.0082 & 0.6861 $\pm$ 0.0092 \\
    ESMM & 0.7348 $\pm$ 0.0008 & 0.7224 $\pm$ 0.0013 &  0.6893 $\pm$ 0.0033\\
    MMoE & 0.7395 $\pm$ 0.0010 & 0.7171  $\pm$ 0.0006 &  0.6907 $\pm$ 0.0013\\
    PLE &  0.7470 $\pm$ 0.0012 & \textbf{0.7309 $\pm$ 0.0008} & 0.7045 $\pm$ 0.0036\\
    AITM &   0.7372 $\pm$ 0.0006 & 0.7104 $\pm$ 0.0011 & 0.7052 $\pm$ 0.0033\\
    MNCM &  0.7420 $\pm$ 0.0011 & 0.7237 $\pm$ 0.0006 & 0.7042$\pm$ 0.0023\\
    PIMM &  $\textbf{0.7489 $\pm$ 0.0004}^{\ddagger}$ & 0.7296 $\pm$ 0.0013 & $\textbf{0.7118 $\pm$ 0.0035}^{\ddagger}$\\
  \bottomrule
    \end{tabular}
      \begin{tablenotes}
        \footnotesize
        \item[$\ddagger$] indicates statistically significant improvement over the best baseline(p-value < 0.05).
      \end{tablenotes}
\end{threeparttable}
}
\end{table}

\textbf{Offline Results}. Generally, the final task in the sequential task paths is typically the most important for practical applications such as recommendations and marketing. Therefore, we mainly focus on the purchase task and core conversion task for the Ali-CCP dataset and industrial dataset respectively. The experimental results on the public dataset are shown in Table \ref{table1}.  These results illustrate that our proposed PIMM significantly outperforms the various state-of-the-art baseline models in terms of the purchase AUC. In Table \ref{table2}, we present the experimental results on the industrial dataset with three tasks. The PIMM model also achieves significant improvement over the baselines in terms of core conversion AUC while maintaining competitive performance on other tasks. The results demonstrate the effectiveness of our proposed PIMM in MTL with sequential dependence tasks. 

\begin{figure}[h]
  \centering
  \includegraphics[width=0.6\linewidth]{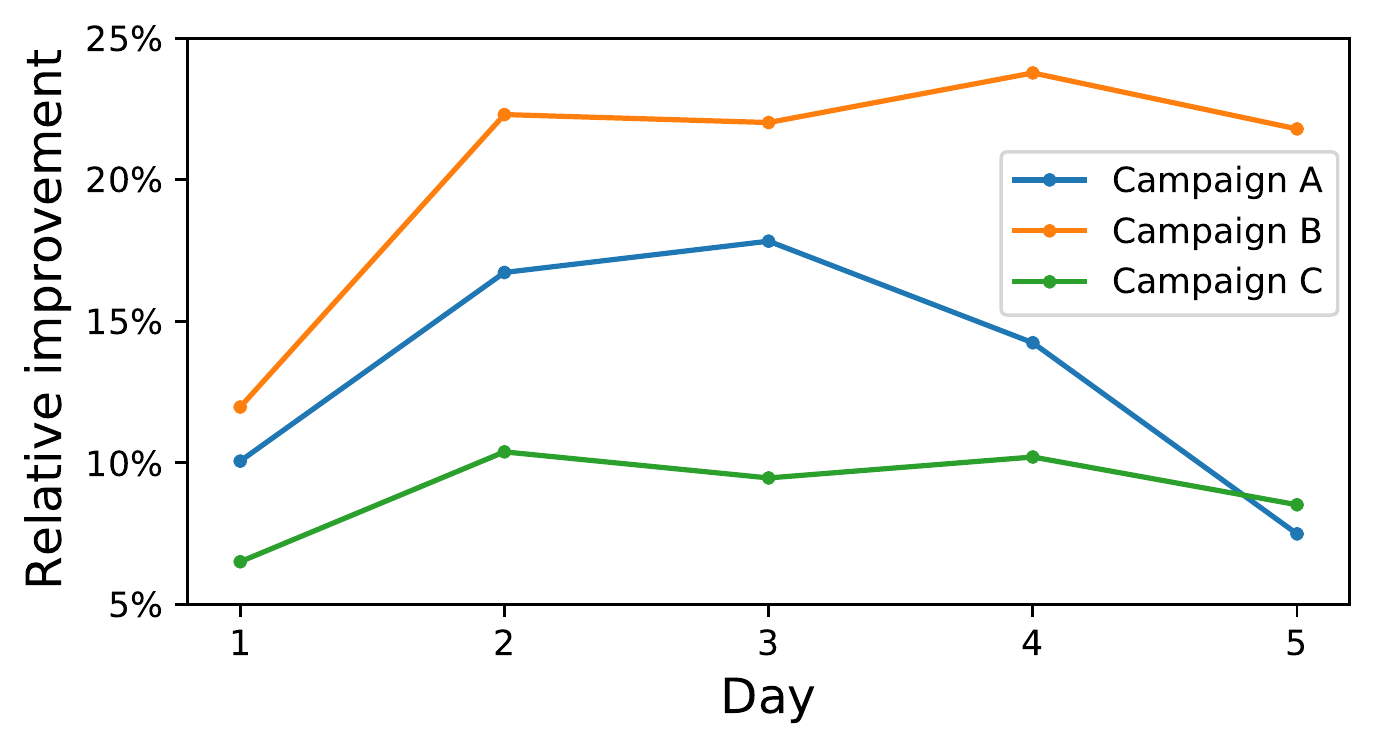}
  \vspace{-0.4cm}
  \caption{The relative improvement of our PIMM compared to baseline in terms of the core conversion customer number.}
  \Description{abt_res}
  \label{abt_res}
\end{figure}

\textbf{Online Results}.  We conduct the online experiment on several real financial customer acquisition campaigns in different periods.  For each campaign, we randomly take 50\% of the traffic as the control group and 50\% of the traffic as the experimental group and ensure two groups are homogeneous. Each model predicts the probability of users' core conversion, and users with the highest potential will be impressed with campaign ads. For a fair comparison, the number of impressions is equal for each group, which means the marketing resources consumed are the same. Figure \ref{abt_res} illustrates the online testing results of three different marketing campaigns. Note that each campaign lasts for 5 days.  We find that PIMM constantly increases the core conversion rate compared with the baseline in all online campaigns. The online results further verify the superiority of the PIMM in modeling the sequential dependence among tasks. 

\section{Conclusion}
In this paper, we propose a novel Prior Information Merged model for multi-task learning with sequential dependence relationships among tasks. The Prior Information Merged module flexibly generates prior information from the ground truth or predictions of the former task. The dynamic sampling mechanism alleviates the error accumulation phenomenon and also guarantees the generalization ability of PIMM.  With the PIM module, PIMM can both leverage the implicit knowledge learned in the task-specific tower of the previous task and the explicit prior knowledge about the dependent task. We conduct extensive offline and online experiments to verify the superiority of PIMM compared with various state-of-the-art multi-task learning models. Besides, the PIM module can be flexibly applied in other advanced model structures.

\bibliographystyle{ACM-Reference-Format}
\bibliography{main}

\end{document}